\documentclass{IEEEtran}  

\IEEEoverridecommandlockouts                              

\makeatletter
\renewcommand{\@IEEEsectpunct}{.\ \,}
\makeatother

\usepackage{graphics} 
\usepackage{epsfig} 
\usepackage{mathptmx} 
\usepackage{times} 
\usepackage{amsmath} 
\usepackage{amssymb}  
\usepackage[]{units}
\usepackage{gensymb}
\usepackage{placeins}
\usepackage{dblfloatfix}  
\usepackage{arydshln}

\title{\LARGE \bf
Rapid Extraction of Respiratory Waveforms from Photoplethysmography: A Deep Encoder Approach}

\author{Harry J. Davies and Danilo P. Mandic\\
(harry.davies14, d.mandic)@imperial.ac.uk
}

\begin{document}

\maketitle
\thispagestyle{empty}
\pagestyle{empty}

\begin{abstract}

Much of the information of breathing is contained within the photoplethysmography (PPG) signal, through changes in venous blood flow, heart rate and stroke volume. We aim to leverage this fact, by employing a novel deep learning framework which is a based on a repurposed convolutional autoencoder. Our model aims to encode all of the relevant respiratory information contained within photoplethysmography waveform, and decode it into a waveform that is similar to a gold standard respiratory reference. The model is employed on two photoplethysmography data sets, namely Capnobase and BIDMC. We show that the model is capable of producing respiratory waveforms that approach the gold standard, while in turn producing state of the art respiratory rate estimates. We also show that when it comes to capturing more advanced respiratory waveform characteristics such as duty cycle, our model is for the most part unsuccessful. A suggested reason for this, in light of a previous study on in-ear PPG, is that the respiratory variations in finger-PPG are far weaker compared with other recording locations. Importantly, our model can perform these waveform estimates in a fraction of a millisecond, giving it the capacity to produce over 6 hours of respiratory waveforms in a single second. Moreover, we attempt to interpret the behaviour of the kernel weights within the model, showing that in part our model intuitively selects different breathing frequencies. The model proposed in this work could help to improve the usefulness of consumer PPG-based wearables for medical applications, where detailed respiratory information is required. 


\end{abstract}


\section{Introduction}

\begin{figure*}[h!]
\centerline{\includegraphics[width=\textwidth]{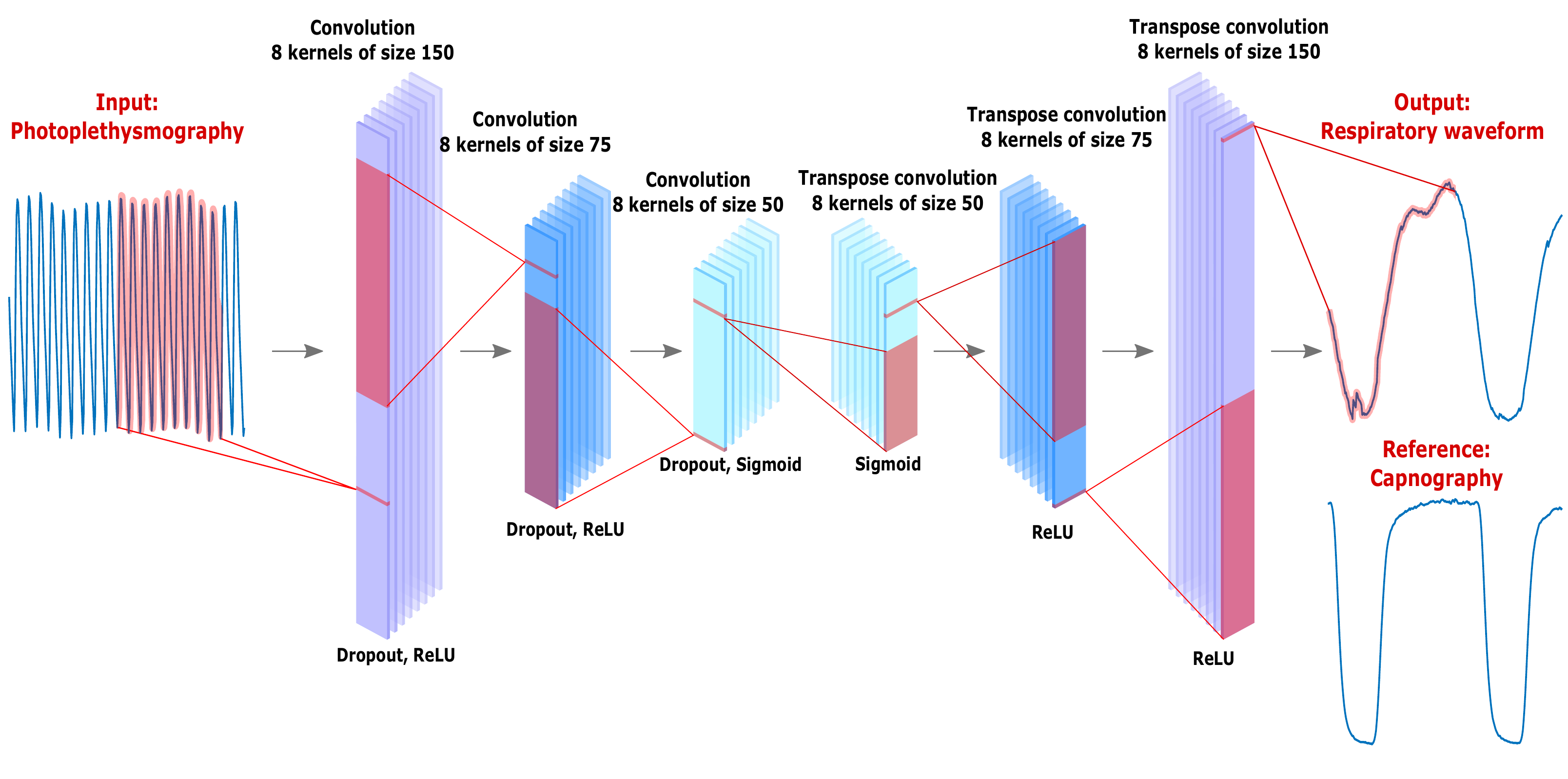}}
\caption{The proposed deep learning architecture. From left to right: Photoplethysmography is passed through 3 one-dimensional convolutional layers each with 8 kernels. The resulting latent space is then passed through 3 transposed one-dimensional convolutional layers which up-sample the latent space into an output respiratory waveform. Under each layer the activation function (rectified linear unit or sigmoid) is labelled, along with the dropout. The output respiratory waveform is depicted next to the reference respiratory waveform, which would be used to calculate mean squared error during training. The waveforms shown in this figure are from a test subject example during leave-one-subject-out cross validation.}
\label{deep_learning_architecture}
\end{figure*}

\IEEEPARstart{W}{earable} health technology promises to revolutionise medicine through avenues of personalised care and long term monitoring of chronic diseases. However, what wearables gain from convenience and quantity of data, they often lose from an available information perspective. A prime example of this is the use photoplethysmography (PPG) for respiratory applications. Whilst the use of PPG has so far demonstrated high accuracy in the monitoring of respiratory rate, much of the respiratory information encoded within PPG is still untapped. To this end, we focus on closing the gap between photoplethysmography and gold standard respiratory monitoring through a multi-layer convolutional encoder-decoder framework that takes PPG as an input and outputs highly accurate respiratory waveforms. We demonstrate that these respiratory waveforms are not just useful for respiratory rate estimation, but are also comparable in terms of mean absolute error with gold standard respiratory measures. Furthermore, this simple deep learning model is computationally cheap to run, making implementation in wearables highly feasible.

\subsection{Photoplethysmography and respiration}

Photoplethysmography (PPG) refers to a non-invasive measurement of blood volume. It operates by emitting a light through the skin and into the underlying tissue, with either one or multiple light emitting diodes, and then measuring the light either transmitted through the tissue (transmittance PPG) or reflected back (reflectance PPG) to one or multiple photodiodes. Blood absorbs light, and therefore when more blood is present less light is reflected back to the photodiode. Thus, PPG can measure changes in blood volume that occur with the pulse and respiration. There is a pressure gradient between the peripheral veins and the heart which results in the return of deoxygenated blood to the heart (known as venous return). Given that the heart lies within the thoracic cavity, and intrathoracic pressure must change to allow for the flow of air in and out of the lungs, respiration therefore modulates this pressure gradient and venous return. When we breathe out, intrathoracic pressure increases to force air out of the lungs and this in turn reduces the pressure gradient between the heart and peripheral veins. A smaller pressure gradient means less venous return to the heart, and an increase in blood volume at the site of the PPG sensor. Moreover, as the lungs constrict more blood flows from the pulmonary veins into the left atrium, then resulting in an increased stroke volume and thus an increased pulse amplitude at the site of the PPG sensor. A larger pulse amplitude with expiration is also accompanied by a slower pulse rate. All three of these changes are reversed during inspiration \cite{Meredith2012}.

To recover respiratory information from PPG, techniques usually rely on the extraction of one or multiple of the three respiratory modes (intensity variations, pulse amplitude variations and pulse interval variations). Common methods of finding respiratory rate include finding the spectral peak corresponding to respiration in the PPG signal \cite{Shelley2006} and band-pass filtering to detect respiratory peaks in the time domain \cite{Nilsson2000}. Photoplethysmography can also be decomposed into physically meaningful frequency components (intrinsic mode functions) through the use of empirical mode decomposition (EMD) or its multivariate equivalent (MEMD) \cite{Rehman2010}. This has been combined with methods such as principal component analysis to select the respiratory modes to give a more accurate respiratory rate estimate \cite{Motin2019}. It has also been shown that respiratory intrinsic mode functions are detailed enough to ascertain the ratio between the duration of inspiration and expiration, allowing for the classification of chronic obstructive pulmonary disease (COPD) \cite{Davies2022}. The methods above are commonly applied to all of the three respiratory variations, and the resulting respiratory rate estimations can then be fused into a final more accurate respiratory rate estimation \cite{Charlton2018}.

\subsection{Capnography}

Capnography refers to the measurment of CO\textsubscript{2} concentrations over time during breathing \cite{Siobal2016}. It is primarily measured from patients who are intubated during scenarios such as surgery. When a patient breathes in, the concentration of CO\textsubscript{2} measured is negligible, whereas when a patient breathes out the concentration of CO\textsubscript{2} is far higher as a result of metabolic processes. In general, this results in a waveform similar to a square wave that acts as a gold standard measure of breathing.

There are certain cases where capnography is atypical, such as a drop in the amplitude of end tidal CO\textsubscript{2} (possibly indicative of a heart failure or pulmonary embolism \cite{Aminiahidashti2018}). Another common example is the where the typical square wave pattern becomes a "shark fin", and this indicates severe airway obstruction such as is the case with COPD. For the purposes of this work we consider how much of the typical capnography respiratory waveform is encoded within the photoplethysmogram, and not emergency medicine scenarios with changes in capnogram amplitude. 

\subsection{Autoencoders for biosignals}

Typically, autoencoders are employed as an unsupervised machine learning technique to learn an efficient coding of an input. The autoencoder operates by compressing an input into a latent space, which should represent information that is fundamental to the input signal. This latent space is then upsampled back into an output, which in typical applications should be the same as the input. These efficient codings of the input that exist in the latent space can then be used directly for classification \cite{gogna2017}, or the output itself can be used for classification \cite{Kang2022}. A typical extension of this principle is for denoising, whereby the input is corrupted by noise and the autoencoder is tasked to match the output with a non-corrupted version of the input. Examples include the denoising of medical images \cite{Gondara2016} and the denoising of biosignals such as the electrocardiogram (ECG) \cite{Chiang2019}.

Fundamentally, the autoencoder structure works by learning a compressed principal relationship between the input and the output. When it comes to PPG, there is clearly respiratory information encoded within the signal. We therefore hypothesize that by using the encoder-decoder structure with photoplethysmography as an input, and training the output against a gold standard respiratory waveform in the form of capnography, it is likely that we can condense and extract all of the relevant respiratory information that is available in PPG.

\section{Methods}

\subsection{Datasets}

\begin{figure*}[h!]
\centerline{\includegraphics[width=0.9\textwidth]{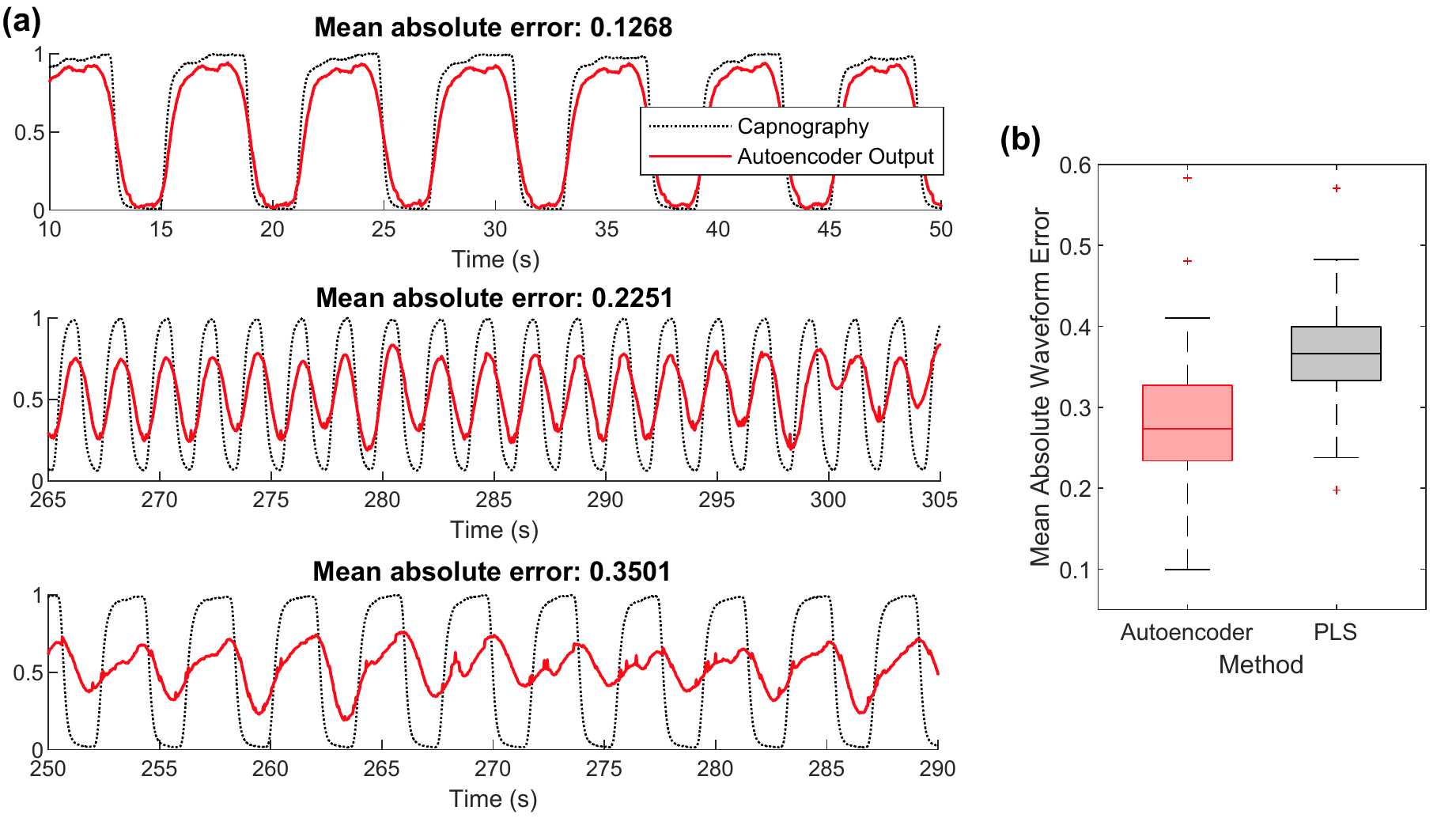}}
\caption{Exemplar output respiratory waveforms and mean absolute error resulting from the proposed model. (a) Output waveforms (red - solid) and gold standard capnography waveforms (black - dotted) for three different scenarios (from top to bottom) of near perfect reconstruction, typical reconstruction and poor reconstruction. (b) Boxplots of test mean absolute waveform error across all 42 subjects of the Capnobase data set, for the proposed deep learning framework (red) and Partial Least Squares regression (grey).}
\label{waveforms_MAE}
\end{figure*}

For the initial leave-one-subject-out cross-validation of our model, we use the Capnobase ``respiratory benchmark" dataset \cite{Walter2013}. This data consists of 42 subjects (29 children and 13 adults) recorded for 8 minutes each during either spontaneous or controlled breathing. The data contains simultaneously recorded finger-based photoplethysmography and capnography, where the capnography serves as a gold standard respiratory reference. The Capnobase encompasses a broad range of respiratory rates, from below 10 breaths per minute to above 40 breaths per minute. The abnormal abundance of higher respiratory frequencies is likely due to a majority of pediatric patients, as children tend to have higher resting respiratory rates. Photoplethysmography can lose respiratory information at higher respiration rates given the low-pass transfer function between breathing and PPG \cite{Johansson1999}. Therefore, having an abundance of higher frequency respiration data for testing, such as is the case in the Capnobase data set, is crucial to ensure an algorithms robustness. Capnobase also contains numerous examples of artefacts where either the capnography or PPG signal is corrupted. It is common for studies to remove epochs of Capnobase recordings that contain artefacts \cite{Rajat2019} \cite{Motin2018}, but in this study we choose to leave these recordings in our analysis for a more realistic view of performance.

For further validation of the model we test on the BIDMC PPG and respiration data set \cite{Pimentel2016} \cite{Goldberger2000}, which contains signals from the MIMIC-II data set. Performance of respiratory rate estimation algorithms on the BIDMC data set is generally worse than on the Capnobase data set, given the strong presence of arterial blood pressure variations at approximately $0.12Hz$ which are difficult to distinguish from respiratory variations \cite{Pimentel2016}. Respiration in the BIDMC data set is derived from two of the electrocardiogram leads being repurposed for impedance pneumography, which tracks the movement of the chest during respiration. Given that this is different in structure and delay to capnography, we cannot directly compare our predicted respiratory waveforms that are trained on capnography to the BIDMC respiratory waveforms in terms of mean absolute error. However, we can compare both waveforms visually and examine the frequency content for respiratory rate estimation. As with the Capnobase data set, we use all of the data in our analysis and do not exclude artefacts. The BIDMC data is resampled to 30Hz, to match the sampling rate of the downsampled Capnobase data.

\subsection{Deep learning model}

A one-dimensional convolutional encoder-decoder structure was implemented in Pytorch \cite{pytorch_cite}. An overview of the model structure is shown in Fig.~\ref{deep_learning_architecture}. It consisted of 3 convolution "encoder" layers with 8 kernels each of sizes 150, 75 and 50 samples, which was then mirrored with 3 upsampling transpose convolution "decoder" layers. Both the PPG input and the output respiratory waveform estimate were 288 samples in length, corresponding to 9.6 seconds of data. The 9.6 seconds was simply a result of dividing the overall 480 seconds of data for each subject into 50 equal segments. Whilst this was somewhat arbitrary, it was important to have a window length that contained at least one full respiratory cycle. A rectified linear activation function was applied to the first two layers, with a sigmoid activation function being applied to the inner-most layers. A dropout of 0.5 was applied to the encoder layers. The input and output of the first layer were both padded by 20, and the output of the second layer was padded by 10. This was again mirrored in the transpose convolution layers within the decoder.

The model was trained by minimising the mean square error between the output and the gold standard capnography. Adam optimisation was implemented with a learning rate of $1\times 10^{-3}$. 

\subsection{Model evaluation}

\begin{figure}[h]
\centerline{\includegraphics[width=0.5\textwidth]{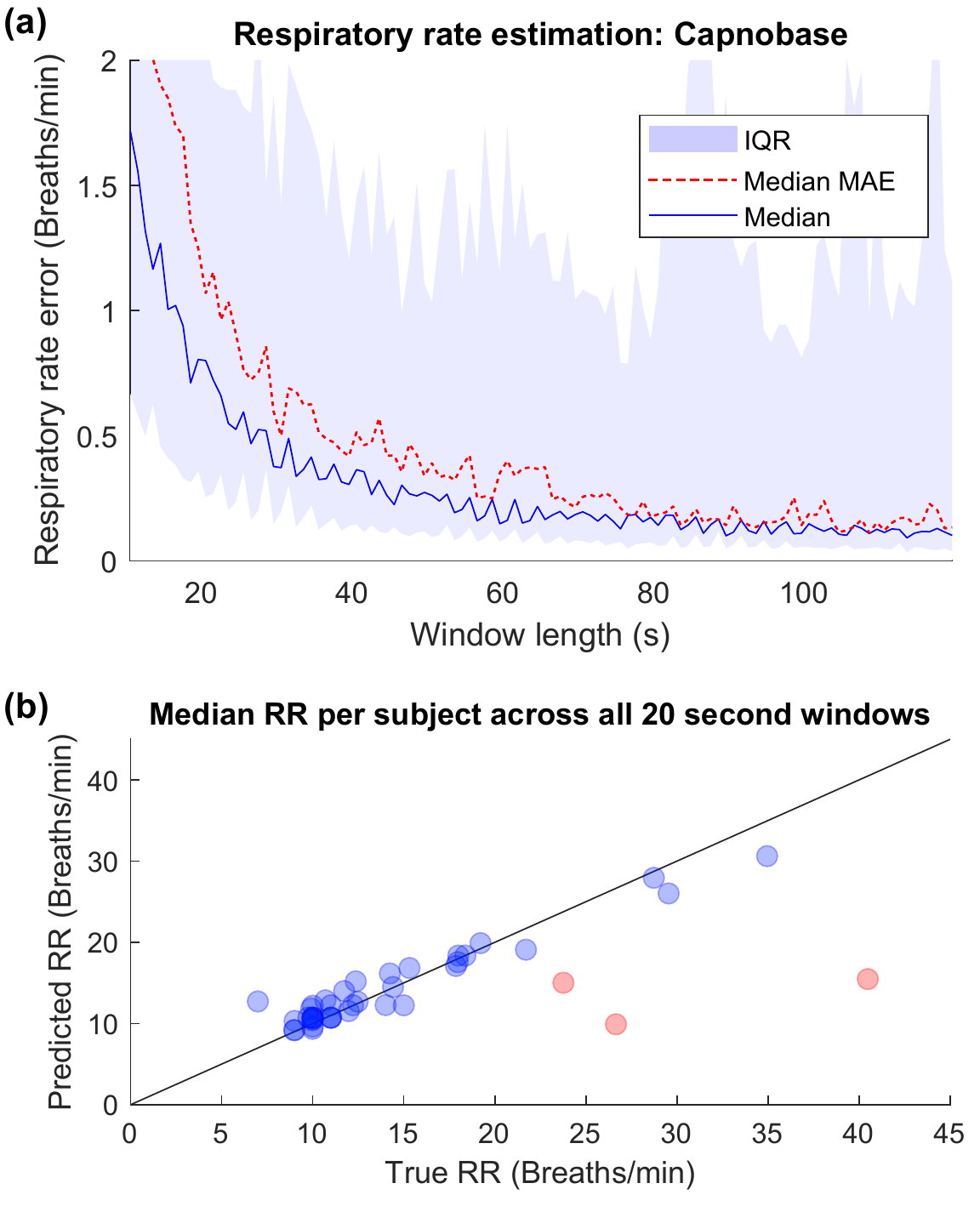}}
\caption{Respiratory rate estimation results for the Capnobase data set. (a) Respiratory rate estimation error across all window lengths, with the median across all windows shown in blue, and the median mean absolute error across all subjects shown in red. (b) Scatter plot of the predicted median respiratory rate against true median respiratory rate for each subject, calculated using all 20 second estimation windows for a given subject. Three of the 42 subjects where the model performed poorly are highlighted in red.}
\label{resp_rate_error}
\end{figure}

\begin{figure}[t]
\centerline{\includegraphics[width=0.5\textwidth]{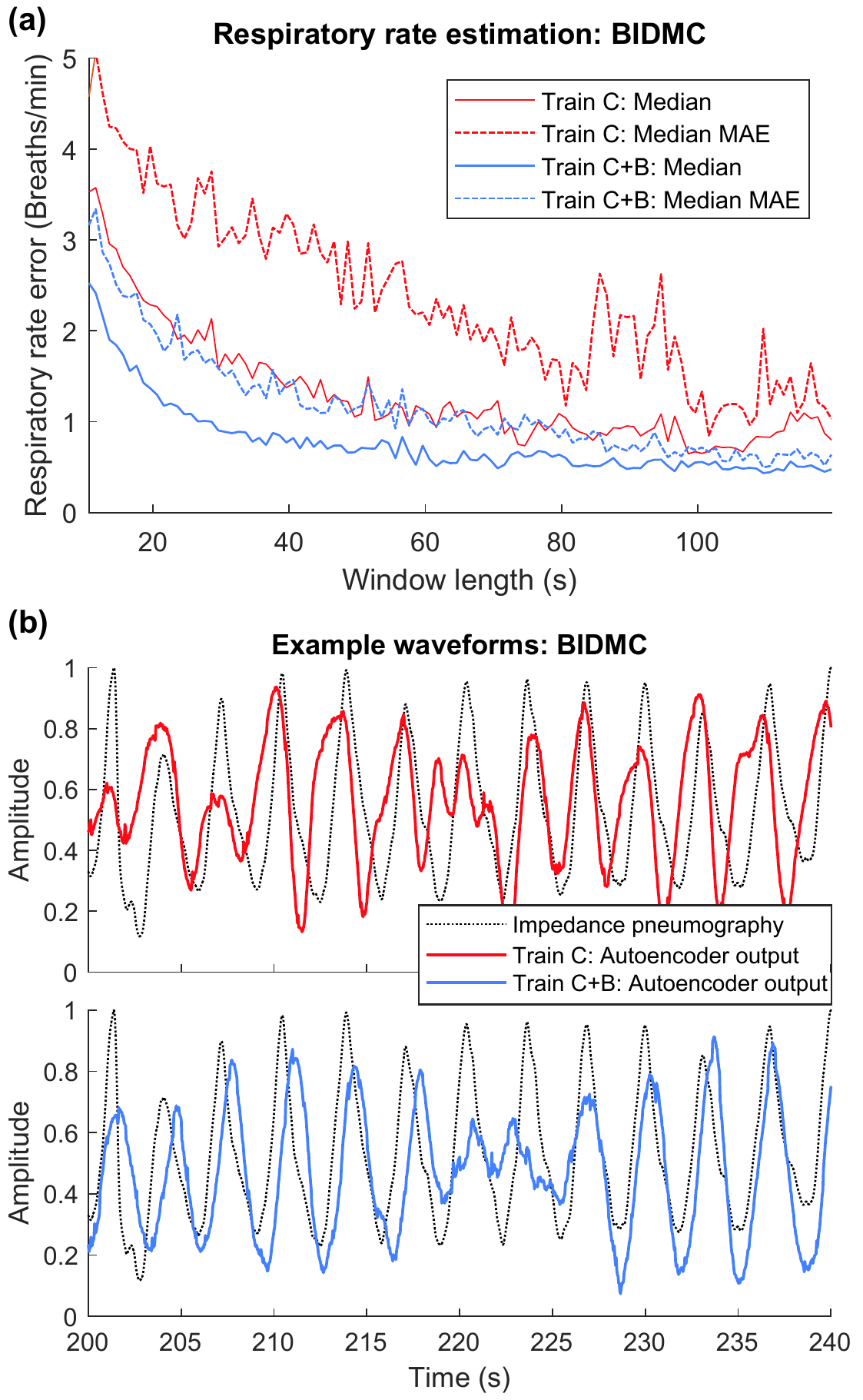}}
\caption{Respiratory rate estimation results and examplar output waveforms for the BIDMC data set. (a) Median absolute error across all windows (solid line) and median mean absolute error across all subjects (dotted line) shown for training on Capnobase and testing on BIDMC (red) and training on Capnobase and then retraining on BIDMC through leave one subject out cross validation (blue). (b) Exemplar output waveforms shown for training on Capnobase and testing on BIDMC (red) and training on Capnobase and then retraining on BIDMC through leave one subject out cross validation (blue), against the gold standard impedance pneumography (grey - dotted).}
\label{bidmc_overview}
\end{figure}

The model was first tested on the Capnobase data set and leave-one-subject-out cross-validation was implemented. When testing on unseen subjects, instead of subjects' photoplethysmography data being divided into 50 unique segments, it was segmented with a sliding window of 9.6 seconds (288 samples) that was shifted 1 second each time. This resulted in 471 test segments for each subject. The model output segments were then fused and averaged, within a given evaluation window size, to give a smoother respiratory waveform estimate.

The output respiratory waveforms were first evaluated by comparing mean absolute error with the reference capnography waveform. The absolute mean error result was also compared to Partial Least Squares (PLS) regression \cite{Dejong1993}, a standard way of finding projection onto latent spaces, which was trained and tested in exactly the same way using MATLAB. In this case, PLS was implemented with 25 degrees of freedom, which was found to give the optimal mean absolute waveform error on test data. This comparison was important given that we are using the encoder-decoder structure in a similar way to what PLS was designed to accomplish.

For the Capnobase data set, the output waveforms were evaluated for both respiratory rate estimation and for the duty cycle (the ratio between inspiration duration and expiration duration). To estimate the respiratory rate, the fast Fourier transform was calculated and the frequency corresponding to the maximum was power was extracted. This frequency in Hertz was then multiplied by 60 to give respiratory rate in breaths per minute. The respiratory rate estimation was compared to the gold standard respiratory rate, calculated manually in the data set from the capnography waveform. To allow comparison with different models in the literature, both the median absolute respiratory rate error across all windows and the median mean absolute error across all subjects was calculated. For duty cycle estimation, given that the waveforms were trained to an amplitude between 0 and 1, the inspiratory duty cycle was estimated as the number of samples below 0.5 was divided by the total number of samples. This was then multiplied by 100 to give a percentage. The same operation was performed on the gold standard capnography for comparison. Both the mean and median absolute error in duty cycle are presented.

For the BIDMC data set, only respiratory rate estimation was used as an evaluation metric, given that the gold standard respiratory waveform was impedance pneumography and no longer capnography. Two training paradigms were implemented for evaluation. Firstly, the model was trained on the entirety of the Capnobase data set and then tested on the BIDMC dataset. Secondly, the model was trained on the Capnobase dataset, and then retrained on the BIDMC dataset in a leave one subject out fashion. In all cases in this work, the model was evaluated on unseen test subjects.

\section{Results}

\begin{table*}[!ht]
    \centering
    \caption{Respiratory Rate Estimation Results}
    \begin{tabular}{lllllll}
    \hline
        Authors & Dataset & Window Size & Error Method & Reported Error (bpm) & \begin{tabular}{@{}c@{}}Proportion of \\ Data Used\end{tabular} & Method  \\ \hline \hline
        Davies (Proposed) & Capnobase & 30.6s & Median AE & 0.37 (0.19-1.42) & 100\% & Encoder-decoder with FFT \\
        Khreis 2019 \cite{Khreis2019new} & Capnobase & 32s & Median AE & 0.5 (0.2-1.1) & 100\% & Kalman Smoothing  \\
        Pimental 2016 \cite{Pimentel2016} & Capnobase & 32s & Median Mean AE & 1.5(0.3-3.3) & 92\% & Spectral fusion \\ \hline
        Davies (Proposed) & BIDMC & 30.6s & Median AE & 0.89 (0.36-3.05) & 100\% & Encoder-decoder with FFT \\
        Aqajari 2021 \cite{Aqajari2021} & BIDMC & 30s & Mean AE & 1.9 +/- 0.3 & 100\% & Generative adversarial networks \\
        Pimental 2016 \cite{Pimentel2016} & BIDMC & 32s & Median Mean AE & 4.0 (1.8-5.5) & 94\% & Spectral fusion \\ \hline
    \end{tabular}
    \label{results_table}
\end{table*}

\begin{figure}[t]
\centerline{\includegraphics[width=0.4\textwidth]{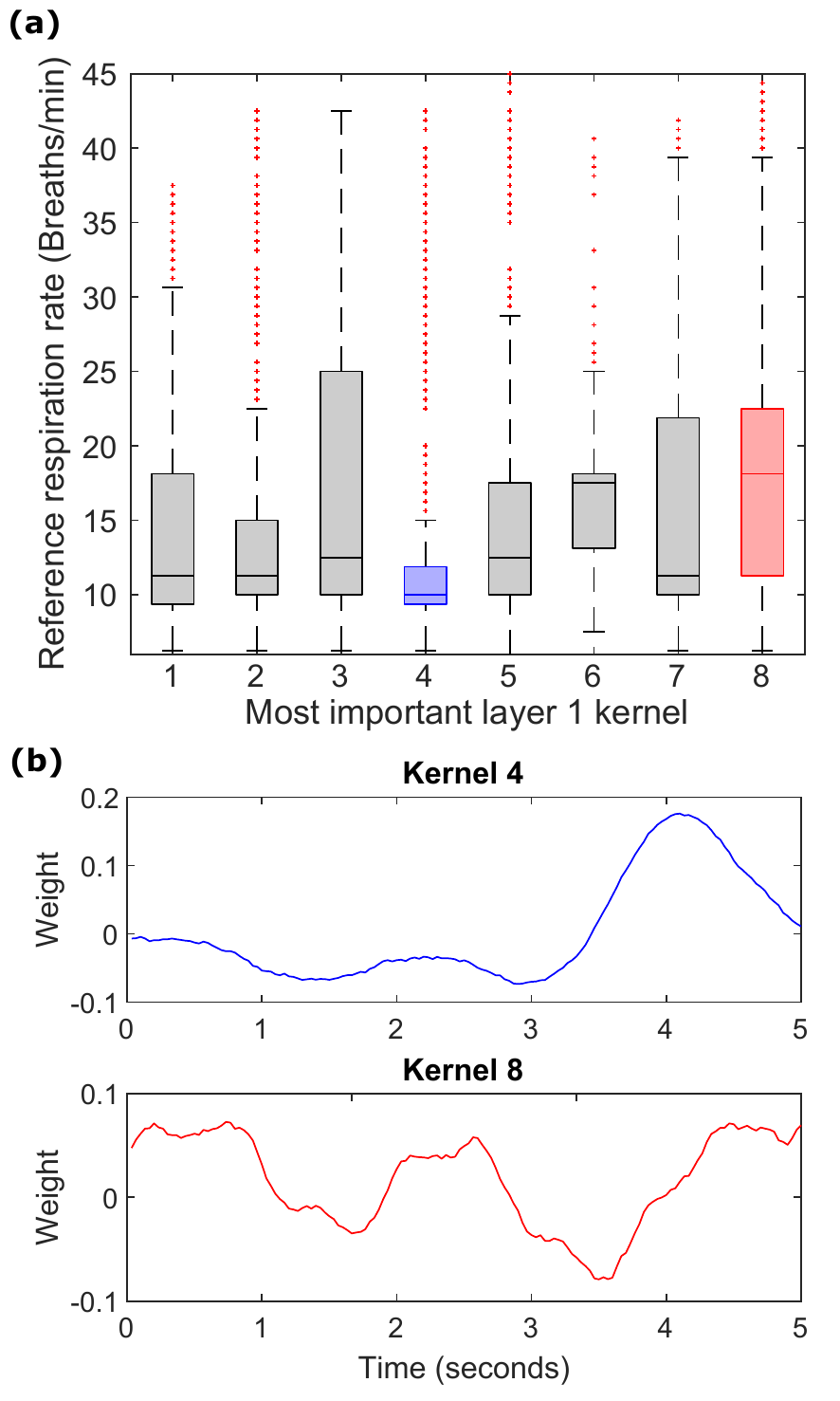}}
\caption{Interpretability of kernel weights according to respiratory rate. (a) Boxplots of the reference respiratory rates for which a given layer 1 kernel produces the largest value in the latent space. The calculation of this is explained in depth in the “Model Interpretability" subsection of the Discussion. (b) Kernel weights for kernel 4 (blue) which has a strong affinity for low frequencies, and kernel 8 (red) which has a strong affinity for higher respiratory rates. Kernels weights in these plots are firstly smoothed with a moving average filter of 1 second in length.}
\label{kernel_boxplot}
\end{figure}

\subsection{Waveform estimation}

Examplar test output waveforms are plotted against the gold standard capnography for three examples of good (MAE of 0.13), typical (MAE of 0.23) and poor (MAE of 0.35) in Fig\ref{waveforms_MAE}(a). It can be seen that even when the difference between waveforms is large in the example with an MAE of 0.35, the basic frequency components are captured. The median mean absolute waveform error of the model outputs across unseen test subjects was 0.27, with an interquartile range of 0.23 to 0.33. This compares to a median MAE of 0.37 and interquartile range of 0.33 to 0.40 for Partial Least Squares regression, as shown in Fig.~\ref{waveforms_MAE}(b).

\subsection{Respiratory rate estimation}

For comparability with other attempts at respiratory rate (RR) estimation in the literature, RR estimation error is presented in both median absolute error (mAE) across all windows, and median mean absolute error (mMAE) across all subjects. For windows of 30.6 seconds, the mAE was 0.37 breaths per minute, and the mMAE was 0.50 breaths per minute. For 60.6 second windows, these results improved slightly with a mAE of 0.16 breaths per minute and a mMAE of 0.40 breaths per minute. The MAE and mMAE are plotted for all time windows from 10 seconds to 120 seconds in Fig.~\ref{resp_rate_error}(a). In Fig.~\ref{resp_rate_error}(b), the median mean error is plotted for each subject across 20.6 second estimation windows. It can be seen that the estimation is broadly accurate across the full range of respiration frequencies with no bias, with estimation being poor in 3 out of the 42 subjects (highlighted in red).

\subsection{Respiratory timing}


When comparing the estimated time spent breathing in, as a proportion of overall breathing cycle, between the output of the deep learning model and the gold standard capnography, there was median Pearson's correlation coefficient of 0.16 (P. = 1.8x10-3). This weak positive correlation indicates that for the most part, respiratory waveform estimations from the finger photoplethysmography failed to fully capture the timing ratio between inspiration and expiration.

\subsection{Validation on BIDMC}

When training the model on the Capnobase dataset, and testing on the unseen BIDMC dataset, respiratory rate estimation was broadly accurate with an MAE and mMAE of 1.74 and 3.02 breaths per minute at 30.6 seconds respectively, and an MAE and mMAE of 1.08 and 2.06 breaths per minute respectively, using 60.6 second windows. When training the model on the Capnobase dataset, and then retraining on the BIDMC dataset, these results were improved with an MAE and mMAE of 0.89, 1.50 and 0.58, 1.01 breaths per minute with 30.6 second windows and 60.6 second windows respectively. The respiratory rate estimate error is shown for all windows between 10 seconds and 120 seconds in Fig.~\ref{bidmc_overview}(a). Moreover, it is shown in Fig.~\ref{bidmc_overview}(b) that with a typical example waveform that with no training on BIDMC, the vast majority of the respiratory information is still captured. Retraining on BIDMC is shown to have the effect of changing the output waveform morphology to resemble that of the impedance pneumography which was in this case used as a gold standard, rather than capnography.

\section{Discussion}

For the reconstruction of capnography waveforms, the encoder-decoder structure presented here outperforms the standard Partial Least Squares regression. It is shown that the model succeeds in near perfect reconstruction of amplitude, phase and respiratory timing in only a small proportion of subjects. In the majority of cases the amplitude information is not captured completely, leading to mean absolute errors greater than 0.23 in three quarters of the test subjects. It can, however, be seen that in waveforms where amplitude information is not captured, resulting in a higher mean absolute waveform error, the dominant frequency information of the respiratory waveform is still effectively captured.

This is further reflected in the capability of the model for accurate estimation of respiratory rate, obtained simply by taking the fast Fourier transform of the output waveforms. Table~\ref{results_table} shows that when compared to other respiratory rate estimation algorithms, the waveforms produced by our model produce similar accuracy to the state of the art in both the Capnobase and BIDMC data sets. When forming these comparisons, care was taken to include recent examples which used all of the data in a given set, as well an algorithm that had been employed on both data sets. Comparisons with algorithms implemented on the BIDMC data set was particularly important, given the difficulty that it presents to respiratory rate estimation through the presence of dominant low frequency non-respiration related variations in the PPG \cite{Pimentel2016}.

Respiratory waveform analysis that goes beyond respiratory rate, such as examining the ratio in timing between breathing in and breathing out (duty cycle), can be important for distinguishing respiratory diseases \cite{TOBIN1983286}. Successful estimation of the inspiratory duty cycle has previously been accomplished from in-ear PPG for the purposes of classifying chronic obstructive pulmonary disease \cite{Davies2022}. Here we show that with the combination of the method presented in this work and the Capnobase data set, extraction of accurate duty cycle was largely unsuccessful, achieving only a weak positive correlation between the true duty cycle and our estimation. The low-pass filter transfer function between breathing and PPG \cite{Johansson1999} generally results in a loss of duty cycle detail \cite{Davies2022}. Respiratory variations are generally weaker from finger PPG than they are from the ear-PPG \cite{Davies2022} \cite{Budidha2018}, which could help to explain the lack of success of duty cycle estimation in this case. 

It is important to note that other capnography waveform differences such as the saw tooth pattern, associated with obstructive breathing disorders, and rare CO\textsubscript{2} amplitude changes were not captured by this model. In the case of specific absolute values of CO\textsubscript{2} concentration in breath, it is fair to assume that this information would not be encoded within the PPG signal. In the case of the saw-tooth waveform however, it is feasible that relevant information is contained within photoplethysmography, and therefore in this case it likely was not captured by the model due to a lack of examples in the data set. 

The relatively low complexity of the deep learning model presented in this work resulted in ultra fast implementation speeds. The mean time for the model to compute a single 9.6 second respiratory waveform window is 0.43 milliseconds on an Intel i7-1165G7 processor (2.8GHz), and therefore over 6 hours worth of respiratory waveforms can be generated in under 1 second. This compares to 5 minutes of data per second when using Kalman smoothing of different frequency estimates (2.4GHz) \cite{Pimentel2016} and 4 minutes of data per second when using an empirical mode decomposition approach combined with independent component analysis (3.2GHz) \cite{Lei2020}. Practically, the speed of our proposed method would allow for easy implementation on small form factor microcontrollers in wearable devices.

\subsection{Model Interpretability}

Especially for health care applications, it is important to attempt to dispel the black box of deep learning methods. Given the relatively low complexity of this model, and convolutional kernels that are comparable in length to the input, we can start to infer what the model is looking for in the input data. To do this, trained weights for the first 3 layers of the model were extracted, and for all inputs windows of photoplethysmography from the Capnobase dataset the latent space was calculated. For each latent space transformation, the maximum was calculated and this result was traced back to the layer 1 kernel that resulted in said maximum. This allows us to infer the most important layer 1 kernel weights for a given input. The most important layer 1 kernels were then plotted against the corresponding reference respiratory rate for every input, resulting in the boxplots shown in Fig.~\ref{kernel_boxplot}(a). It can be seen that in some part these kernels are discriminating the different inputs based on frequency. This deduction is further reinforced when we visualise the smoothed weights of kernel 4 (triggering predominantly at low frequency respiration with a median reference respiratory rate of 10 breaths per minute) and kernel 8 (triggering predominantly at higher frequencies with a median reference RR of 18 breaths per minute) in Fig.~\ref{kernel_boxplot}(b). The weights of kernel 4 correspond to a low frequency pattern of respiration, whereas the weights of kernel 8 contain higher frequency information. This result is important as it shows that the input model is in part extracting frequency based information from the input, in a similar way to performing a discrete cosine transform. This knowledge helps us to move from a black box interpretation of the model towards a grey box. 

\section{Conclusion}

We have introduced a novel low complexity deep learning framework for the extraction of respiratory information from photoplethysmography. The proposed model has been shown to produce accurate respiratory waveforms and state of the art respiratory rate estimates. We have demonstrated this effectiveness on two different data sets, Capnobase and BIDMC, highlighting its strong ability to generalise. Importantly, given the relative simplicity of our model, it has been shown to be capable of transforming several hours of photoplethysmography data into corresponding respiratory waveforms in a single second. The framework presented in this work helps to bridge the gap between photoplethysmography and gold standard respiratory references such as capnography and impedance pneumography. We believe that our model can therefore improve the usefulness of consumer PPG-based wearables in medical applications.

\section*{Acknowledgment}
This work was supported by the Racing Foundation grant 285/2018, MURI/EPSRC grant EP/P008461, and the Dementia Research Institute at Imperial College London.

\FloatBarrier
\bibliographystyle{IEEEtran}
\bibliography{IEEEabrv,capnography_ppg_autoencoder}

\end{document}